# Reduction of 1/$f$ Noise in Graphene after Electron-Beam Irradiation

Md. Zahid Hossain[1], Sergey Rumyantsev[2,3], Michael S. Shur[2] and Alexander A. Balandin[1,4,*]

[1]Nano-Device Laboratory, Department of Electrical Engineering, Bourns College of Engineering, University of California – Riverside, Riverside, California 92521 USA

[2]Center for Integrated Electronics and Department of Electrical, Computer and Systems Engineering, Rensselaer Polytechnic Institute, Troy, New York 12180 USA

[3]Ioffe Physical-Technical Institute, The Russian Academy of Sciences, St. Petersburg, 194021 Russia

[4]Materials Science and Engineering Program, University of California – Riverside, Riverside, California 92521 USA

[*]Corresponding author (AAB):  balandin@ee.ucr.edu

**Abstract**

We investigated the effect of the electron-beam irradiation on the level of the low-frequency $1/f$ noise in graphene devices. It was found that $1/f$ noise in graphene reveals an *anomalous* characteristic − it reduces with increasing concentration of defects induced by irradiation. The increased amount of structural disorder in graphene under irradiation was verified with micro-Raman spectroscopy. The bombardment of graphene devices with 20-keV electrons reduced the noise spectral density, $S_I/I^2$ ($I$ is the source-drain current) by an order-of magnitude at the radiation dose of $10^4$ $\mu C/cm^2$. Our theoretical considerations suggest that the observed noise reduction after irradiation can be more readily explained if the mechanism of $1/f$ noise in graphene is related to the electron-mobility fluctuations. The obtained results are important for the proposed graphene applications in analog, mixed-signal and radio-frequency systems, integrated circuit interconnects and sensors.

The level of the flicker $1/f$ noise [1] is one of the key metrics that each new material has to pass before it can be used for practical devices ($f$ is the frequency) [2]. Graphene [3] has shown a great potential for applications in high-frequency communications [4-5], analog circuits [6] and sensors [7-8]. The envisioned applications require a low level of $1/f$ noise, which contributes to the phase-noise of communication systems [2] and limits the sensor sensitivity [7]. Despite significant research efforts [9-15] there is still no conventionally accepted model for physical mechanisms behind $1/f$ noise in graphene. Correspondingly, no comprehensive methods for $1/f$ noise suppression in graphene devices have been developed.

In this Letter we show that $1/f$ noise in graphene reveals an *anomalous* characteristic – it reduces with increasing concentration of defects induced by irradiation. We found that bombardment of graphene devices with 20-keV electrons can reduce the noise spectral density, $S_I/I^2$ ($I$ is the source-drain current) by an order-of magnitude at the radiation dose (RD) of $10^4$ μC/cm$^2$. Our theoretical analysis suggests that the observed noise suppression after introduction of defects can be explained if the mechanism of $1/f$ noise in graphene is related to the electron-mobility fluctuations rather than to the carrier-density fluctuations. Apart from contributing to understanding the physics behind $1/f$ noise in graphene our results can possibly offer a practical method for noise reduction in various graphene devices.

Graphene revealed a number of unique electronic [3-7], optical [7] and thermal [16] properties, which led to proposals of different graphene-based devices. The high-frequency communication and analog circuit applications are expected to capitalize on high electron mobility and saturation velocity of graphene [7]. The ultimate surface-to-volume ratio and the Fermi-energy tuning capability make graphene excellent material for detectors and sensors with demonstrated single-molecule sensitivity [7] and selective gas sensing [8]. However, like with any other new material systems, graphene has to meet the stringent requirements for the low-frequency $1/f$ noise level [2]. For example, recent development of GaN technology for radio-frequency and optical communications required substantial decrease of the $1/f$ noise in this material [2, 17].

A number of research groups have studied low-frequency noise in graphene devices [9-15]. The noise spectral density, $S_I/I^2$, follows $1/f$ law but reveals the gate-bias dependence different from that in semiconductor field-effect transistors [9-15]. The noise level was reported to be smaller in bilayer [9] or graded-thickness [14] few-layer graphene than in single-layer graphene. In many cases, the interpretations of measured results were different. As of today, there is no commonly accepted model of $1/f$ noise in graphene. The fundamental $1/f$ noise mechanism in graphene – mobility fluctuations vs. carrier number fluctuations – is still not known. The latter provided a strong motivation for the present study. The irradiation experiments were crucial for gaining understanding $1/f$ noise in conventional materials and metal-oxide-semiconductor field-effect transistors [18-20]. Investigation of the effects produced by the electron beams on $1/f$ noise in graphene can elucidate the mechanisms of $1/f$ noise and answer a question of graphene's prospects for radiation-hard applications.

Graphene is relatively susceptible to the electron and ion bombardment owing to its single-atom thickness [21-23]. Electron irradiation can introduce different types of defects in graphene depending on the beam energy and local environment, e.g. presence of organic contaminants. For this study we selected the electron energy of 20 keV in order to exclude the severe knock-on damage to the graphene crystal lattice, which starts at ~50 keV [23]. The back-gated graphene devices were fabricated using mechanical exfoliation [5] and standard lithographic fabrication techniques [10, 14]. The number of atomic planes and absence of defects in pristine graphene have been checked with Raman spectroscopy. The devices were subjected to the electron-beam irradiation in the scanning-electron microscopy (SEM) chamber of the electron-beam lithography (EBL) system under high-vacuum conditions. The following testing protocol was applied. The Raman spectrum and $1/f$ noise were measured for all devices before irradiation. The graphene channels were then exposed to the first RD followed by the Raman spectroscopy and noise measurements. The process was repeated several times until the total RD reached $10^5 \mu C/cm^2$.

Figure 1(a) shows a SEM image of a graphene channel with four metal contacts while Figure 1(b) illustrates the area of graphene between two contacts subjected to irradiation. The Raman spectrum of graphene before and after irradiation with RD=$10^4$ $\mu C/cm^2$ is presented in Figure 1(c). One can see the appearance of the strong disorder $D$ and $D'$ peaks after irradiation indicating that electron bombardment introduced defects to graphene [22]. The strength of the irradiation effects can be deduced from the intensity ratio, $I(D)/I(G)$, presented in Figure 1(d). $I(D)/I(G)$ increases monotonically with RD in the range of interest. Defect introduction results in corresponding asymmetric broadening of $2D$ band, which can also be used to monitor defect introduction.

The electrical conduction properties of graphene under irradiation evolved as we expected. The electron bombardment led to a shift in the Dirac point position and decrease in mobility, $\mu$, with the corresponding increase in the source-drain resistance, $R_{SD}$. The initial position of the Dirac point ($V_D \sim 10$ V) is typical for as-fabricated devices owing to the background doping from water moisture or resist residues from lithographic processes. The Dirac point shifted to $\sim 2$ V after exposure to RD=$10^4 \mu$C/cm$^2$ (Figure 2(a)). Such a behavior was observed for most devices, although in a very few cases, we recorded a positive shift of the Dirac point after some irradiation steps. The mobility decreases with increasing RD but remains acceptable from the applications point of view (Figure 2(b)).

The low-frequency noise measurements were performed in the in-house built setup shielded inside a metal enclosure. The details of the measurements were described by us elsewhere [10, 14] (see also Supplementary Information). Figure 3(a) shows the noise amplitude as a function of the gate bias and channel resistance in pristine graphene. Its behavior is in agreement with previous reports indicating that the noise characteristic of our devices were typical for graphene [9 - 15]. Figure 3(b) presents the noise spectral density, $S_V/I^2$, for a graphene device before irradiation and after each irradiation step. $S_V/I^2$ remains of $1/f$ – type before and after irradiation. The most surprising observation from Figure 3(b) is that the noise level in graphene *decreases* after irradiation. The outcome was *reproducible* as proven by repeating the measurements for a large number of devices (>20). Figure 4(a-b) shows that the noise reduces monotonically with the increasing RD for the entire range of negative gate-bias voltages, $V_G$-$V_D$. The same trend was observed for the positive gate bias (see Supplementary Information).

The reduction of 1/*f* noise after irradiation is *highly unusual* and *counterintuitive*. The flicker 1/*f* noise in semiconductors and metals is usually associated with structural defects. Therefore, introduction of defects by electron, ion, gamma or X-ray irradiation normally results in increased levels of 1/*f* noise [2, 18-20]. However, as an exception to the rule, there have been a few reports when the 1/*f* noise decreased as a result of irradiation [24-25]. In all previously reported cases of the 1/*f* noise reduction – a specific mechanism was responsible for the observed effect. Often, it was related to a particular device design [24-25]. In our investigation, we employed the simplest device structure – generic graphene channel with a back-gate for controlling the number of carriers. It appears that the noise reduction phenomenon in graphene is of more *fundamental nature* related to the specifics of the electron transport as discussed below.

There are two basic mechanisms of low-frequency noise in electrical conductors: fluctuations of the number of carriers or fluctuations of the charge-carrier mobility. In semiconductors and transistors, 1/*f* noise usually complies with the classical McWhorter model of the number of carriers fluctuations [26]. In this model, 1/*f* spectrum is a result of superposition the electron trapping and de-trapping events mediated by tunneling to the oxide traps located at difference distances from the channel. Since the tunneling probability depends exponentially on the distance between a trap and the channel there is an exponentially wide distribution of the characteristic times, $\tau$. A wide distribution of $\tau$ results in the 1/*f* overall noise spectral density is written as [27-28]

$$\frac{S_I}{I^2} = \frac{\lambda k T N_t}{f A V n^2}. \qquad (1)$$

Here $N_t$ is the concentration of the traps near the Fermi level responsible for noise, $A$ is the gate area, $n$ is the carriers concentrations and $\lambda$ is the tunneling constant. One can see from Eq. (1) that the only way to explain the noise reduction within the number-of-carriers fluctuation mechanism is to assume the decrease in $N_t$ as result of irradiation. Figure 2(a) shows that the Dirac point shifts to the smaller positive value after irradiation. Therefore for the same gate voltage the Fermi level position is different in the pristine and irradiated samples. Since the maximum contribution to the noise comes from the traps located near the Fermi level, the reduction of noise is not prohibited in the framework of the McWhorter model. However, the overall gate-voltage dependence of noise does not comply with the McWhorter model. While McWhorter model predicts $1/n^2$ dependence of noise, experiments with graphene show a variety of dependences including weak decrease or increase of noise with the gate voltage, i.e. with the concentration $n$ [9-15]. The above scenario also leads not only to decrease of noise at some gate bias but also to increase of the noise at other gate bias, which was not observed experimentally. Therefore we believe that the observed noise reduction owing to the irradiation induced change in the concentration of traps, which contribute to $1/f$ noise, is unlikely or, at least, not dominant mechanism.

In the framework of the mobility-fluctuation model, the noise spectral density of the elemental fluctuation events contributing to $1/f$ noise in any material is given by [27-28]

$$\frac{S_I}{I^2} \propto \frac{N_t^\mu}{V} \frac{\tau \varsigma (1-\varsigma)}{1+(\omega \tau)^2} l_0^2 (\sigma_2 - \sigma_1)^2, \qquad (2)$$

where $N_t^\mu$ is the concentration of the scattering centers of a given type, $l_0$ is the mean free path of the charge carriers, $\zeta$ is the probability for a scattering center to be in the state with the cross-section $\sigma_l$. Integration of Eq. (2) results in the $1/f$ spectrum caused by the mobility fluctuations. One should note that $N_t^\mu$ is not a concentration of the scattering centers, which limit the electron mobility but rather a concentration of the centers contributing to the mobility fluctuations. The number of such centers may increase or stay unchanged as a result of irradiation. Mobility and, correspondingly, $l_0$ ($\sim\mu$) decrease as a result of irradiation (see Figure 2(b)) leading to the reduction in $S_I/I^2$ ($\sim l_0^2$). In graphene, $\mu$ is limited by the long-range Coulomb scattering from charged defects even at room temperature (RT) [29-30], in contrast to semiconductors or metals, where $\mu$ at RT is typically limited by phonons, even if the defect concentration is high. Thus, it follows from our measurements and Eq. (2) that mobility-fluctuation mechanism can explain the noise reduction as a result of irradiation over the entire gate-bias range. Our results also suggest that the defects generated in graphene by irradiation are not the type of defects with the switching ionization because that would yield a normalized $1/f$ noise proportional to the mobility [31].

The noise reduction after irradiation came at the expense of the mobility degradation from the average value of 3000 cm$^2$/Vs to about 1000 cm$^2$/Vs at RT. Irradiating the device with the higher initial mobility ($\mu$=5000 cm$^2$/Vs) with the same RD resulted in $\mu\sim$ 2000 cm$^2$/Vs, consistent with prior mobility studies [32]. This reduction in mobility although substantial does not preclude practical applications. It has been noted that for graphene devices used in nanometer-scale architectures the mobility above $\sim$1000-1500 cm$^2$/Vs is not even needed owing to the onset of the ballistic transport regime when the source-drain distance shrinks to the deep-submicron limit [33].

In conclusion, we discovered that $1/f$ noise in graphene reveals an unusual feature: it reduces upon introduction of defects. Considering two basic mechanisms of $1/f$ noise we established that the observed noise reduction can be more readily explained within the mobility-fluctuation model. We anticipate that other types of irradiation can produce similar effects although the required energies and doses can be different. The described *defect engineering* technique can bring about a conceptual change to the device reliability methods and open up alternative routes to graphene radiation-hard applications in aviation, space and medical fields.

*Acknowledgements:* The work at UCR was supported, in part, by the Semiconductor Research Corporation (SRC) and Defense Advanced Research Project Agency (DARPA) through FCRP Center on Functional Engineered Nano Architectonics (FENA) and by the National Science Foundation (NSF) projects US EECS-1128304, EECS-1124733 and EECS-1102074. The work at RPI was supported by the US NSF under the auspices of I/UCRC "CONNECTION ONE" at RPI and by the NSF EAGER program. SLR acknowledges partial support from the Russian Fund for Basic Research (RFBR) grant 11-02-00013.

**Author Contributions:** A.A.B. envisioned the experiments, coordinated the project, contributed to data analysis and wrote the manuscript; M.Z.H. fabricated graphene devices, performed Raman and noise measurements, contributed to data analysis and manuscript preparation; S.R.

contributed to data analysis and manuscript preparation; M.S.S. contributed to data analysis and manuscript preparation.

**FIGURE CAPTIONS**

**Figure 1: Graphene device irradiation with electron beams.** (a) SEM image of graphene devices with multiple metal contacts. The dark ribbons are graphene channels while the white regions are Ti/Au(10-nm/90-nm) electrodes. The scale bar is 2 μm. (b) Schematic of the irradiation process showing the area exposed to the electron beam. (c) Raman spectrum of graphene before and after irradiation. The single-layer graphene signatures include $G$ peak at ~1584 cm$^{-1}$ and symmetric $2D$ band at ~2692 cm$^{-1}$. The absence of the disorder $D$ peak at ~1350 cm$^{-1}$ proves that graphene is high quality and defect-free before irradiation. Appearance of the disorder $D$ and $D´$ peaks after irradiation indicates that electron bombardment introduced defects to graphene. (d) Intensity ratio $I(D)/I(G)$ as a function of the irradiation dose. The inset shows the normalized $2D$ band at different irradiation doses shifted in energy to the same position for the ease of comparison. Note the asymmetric broadening and skewing toward the lower wave numbers. The full-width at half maximum of $2D$ band before irradiation was ~28 cm$^{-1}$ while after irradiation it increased to ~36 cm$^{-1}$ at the irradiation dose of $5\times10^4$ μC/cm$^2$.

**Figure 2: Irradiation effects on electrical characteristics of graphene.** (a) Source-drain current as a function of the back-gate bias. The position of the Dirac point shifts to a smaller voltage as a result of irradiation. (b) Electron mobility dependence on the irradiation dose for two devices. The mobility values for the pristine devices were in the range from 2500 to 5000 cm$^2$V/s at room temperature. The inset shows a corresponding increase of the graphene channel

resistance after irradiation for one of the devices. The initial areal irradiation dose of 300 $\mu$C/cm$^2$ is comparable to the typical dose of 500 $\mu$C/cm$^2$ in the lithographic process.

**Figure 3: Noise suppression in graphene via electron beam irradiation.** (a) Noise amplitude $A = (1/N)\Sigma_{m=1}^{N} f_m S_{I_m} / I_m^2$ as the function of the gate bias and channel resistance in pristine graphene. Note that the noise amplitude is a metric similar to the normalized noise spectral density, $S_I/I^2$, but involves averaging over several frequencies. The data indicates that our devices reveal noise behavior consistent with previous reports. (b) Noise spectral density, $S_I/I^2$, as a function of frequency for a graphene device shown after each irradiation step. $S_I/I^2$ is normalized to the source-drain current, $I \equiv I_{SD}$. The source-drain DC bias was varied between 10 mV to 30 mV during the noise measurements. Note that the 1/$f$ noise *decreases* monotonically with the increasing irradiation dose. $S_I/I^2$ is more than an *order-of-magnitude* smaller after 5×10$^4$ $\mu$C/cm$^2$ radiation does than that in pristine graphene.

**Figure 4: Mechanism of the noise suppression in graphene.** (a) $S_I/I^2$ as a function of the radiation dose at zero gate bias for three frequencies $f$= 20, 40 and 100 Hz. The arrows indicate the level of 1/$f$ noise before irradiation. (b) $S_I/I^2$ as the function of the gate bias, $V_G$, referenced to the Dirac point, $V_D$, for another graphene device before and after irradiation plotted for $f$=20 Hz. The negative bias corresponds to the hole-transport regime. Note that the noise suppression in graphene via defect engineering works in the entire range of biasing conditions and frequencies pertinent to practical applications.

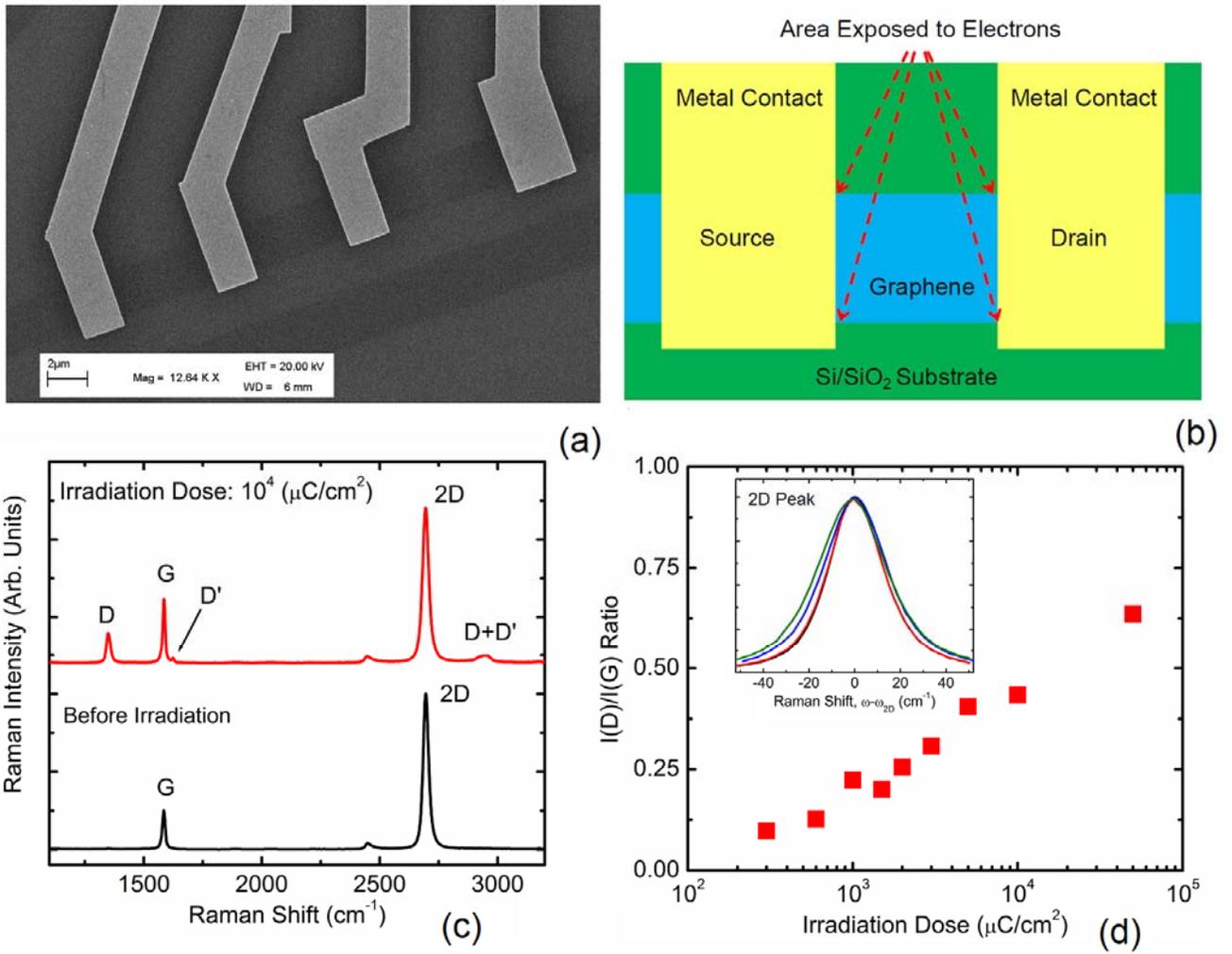

Figure 1

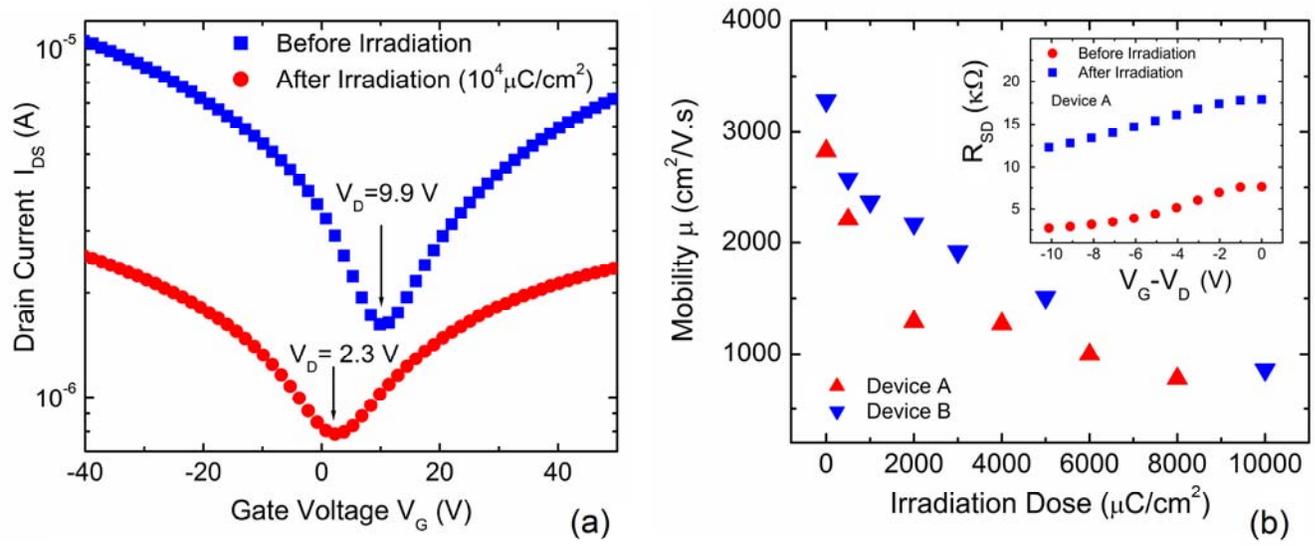

Figure 2

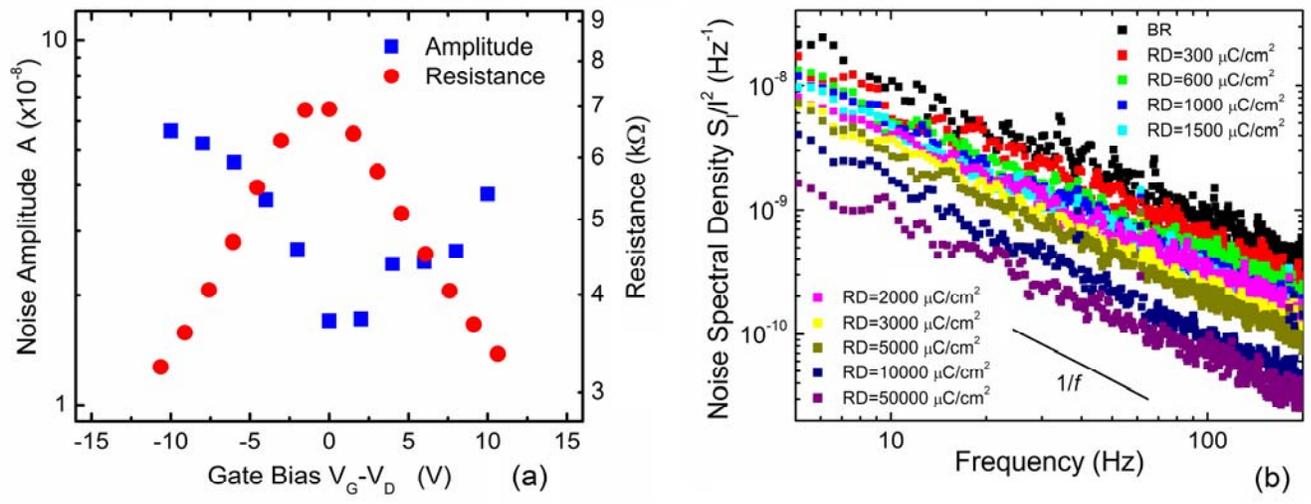

Figure 3

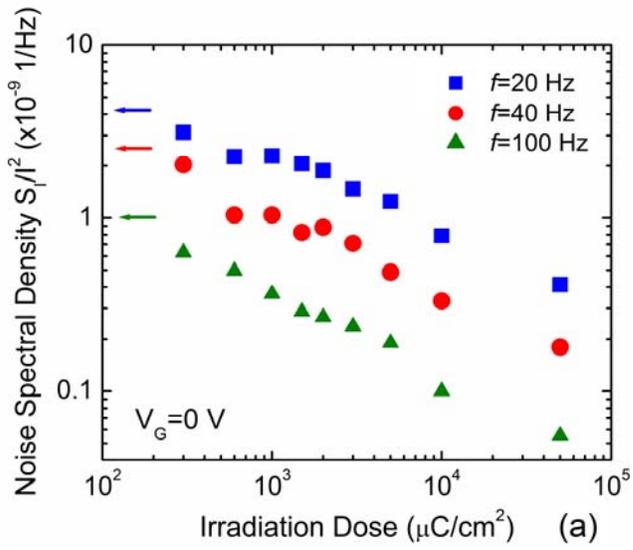

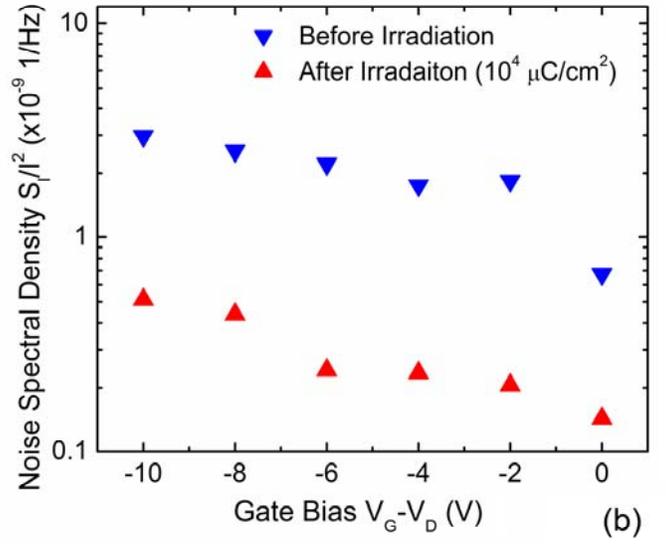

Figure 4